\documentclass[aps,prx,twocolumn,tightenlines,amsmath,amssymb,superscriptaddress, floatfix]{revtex4-1}
\usepackage{graphicx}
\usepackage{amsmath}
\usepackage{amssymb}
\usepackage{epstopdf}
\usepackage{color}
\usepackage{lipsum}
\usepackage[colorlinks=true]{hyperref}
\usepackage{ulem}
\usepackage[dvipsnames]{xcolor}
\usepackage{array}
\usepackage{comment}
\usepackage{braket}
\usepackage{xr}

\begin{document}

\title{Accessing which-path information in the absorption and emission of light by a quantum dot in a Ramsey sequence}%

\date{\today}

\author{I. Maillette de Buy Wenniger}
\email{ilse.maillettedebuywenniger@physics.ox.ac.uk}
\affiliation{Clarendon Laboratory, University of Oxford, Parks Road, Oxford OX1 3PU, United Kingdom.
}%
\affiliation{Université Paris-Saclay, CNRS, Centre de Nanosciences et de Nanotechnologies, 91120, Palaiseau, France.} 

\author{M. Maffei}
\affiliation{Université de Lorraine, CNRS, Laboratoire de Physique et Chimie Théoriques, 54500 Vandoeuvre-les-Nancy, France.}%

\author{S. C. Wein}%
\affiliation{%
Quandela SAS, 91300 Massy, France.
}%

\author{S. P. Prasad}
\affiliation{MajuLab, CNRS-UCA-SU-NUS-NTU International Joint Research Laboratory, 117543 Singapore, Singapore.}
\affiliation{Centre for Quantum Technologies, National University of Singapore, 117543 Singapore, Singapore.}

\author{H. Lam}
\affiliation{Université Paris-Saclay, CNRS, Centre de Nanosciences et de Nanotechnologies, 91120, Palaiseau, France.} 
 
\author{D. Fioretto}
\affiliation{Université Paris-Saclay, CNRS, Centre de Nanosciences et de Nanotechnologies, 91120, Palaiseau, France.} 
\affiliation{%
Quandela SAS, 91300 Massy, France.
}%

\author{A. Lema\^itre}
\affiliation{Université Paris-Saclay, CNRS, Centre de Nanosciences et de Nanotechnologies, 91120, Palaiseau, France.} 

\author{\\ I. Sagnes}
\affiliation{Université Paris-Saclay, CNRS, Centre de Nanosciences et de Nanotechnologies, 91120, Palaiseau, France.} 

\author{C. Ant\'on-Solanas}
\affiliation{Depto. de F\'isica de Materiales, Instituto Nicol\'as Cabrera, Instituto de F\'isica de la
Materia Condensada, Universidad Aut\'onoma de Madrid, 28049 Madrid, Spain.
}%
\author{Pascale Senellart}
\affiliation{Université Paris-Saclay, CNRS, Centre de Nanosciences et de Nanotechnologies, 91120, Palaiseau, France.} 

\author{A. Auff\`eves}
\email{alexia.auffeves@cnrs.fr}
\affiliation{MajuLab, CNRS-UCA-SU-NUS-NTU International Joint Research Laboratory, 117543 Singapore, Singapore.}
\affiliation{Centre for Quantum Technologies, National University of Singapore, 117543 Singapore, Singapore.}
\date{\today}

\begin{abstract}
We quantify which-path information in the absorption and emission of light by a quantum dot along a Ramsey-like sequence. The quantum dot is excited by two successive classical $\pi/2$-pulses with tunable relative phase, yielding the spontaneous release of coherent superpositions of zero- and one-photon Fock states into two successive time bins. Along the sequence, the first time bin extracts information on the quantum dot energy state, behaving as a which-path detector for the Ramsey interferometer. The which-path information increases over time, and is accessed through the reduction of contrast of the Ramsey fringes. After the second pulse, the information still present in the first time bin controls the emission of coherent light into the second time bin, which is measurable through the reduction of the contrast of self-homodyne interference fringes in a Mach-Zehnder interferometer. Both measurements are in remarkable agreement with theoretical predictions. Our results quantitatively illustrate how which-path information and more generally quantum correlations impact light-matter energy exchanges in the quantum realm.
\end{abstract}

\maketitle

\textit{Introduction--} Which-path (WP) information is a central concept in quantum physics. It has provided deep insights into the theory of measurement and its relation to entanglement, complementarity and wave-particle duality~\cite{Wooters-Zurek79, Walther91, Wineland98}. It lies at the core of so-called quantum eraser and delayed choice experiments~\cite{QEraser}. WP information typically appears in interferometric experiments involving two possible evolution paths with tunable relative phase. The resulting interference fringes have maximal visibility if the paths are indistinguishable (wave-like behavior). In contrast, the presence of a WP detector -- a quantum system whose states become correlated with the path taken by the system, playing the role of a quantum meter -- reduces the fringe contrast. It leads to its complete disappearance if the information recorded is complete (particle-like behavior). WP information is quantified by the distinguishability of the quantum meter's states: the larger the distinguishability, the smaller the visibility of the interference~\cite{Englert96,Rempe98}. 

WP information has been measured in optical interferometers inspired by Young's double slit experiment. In such implementations, two laser-driven atoms scatter light and effectively play the role of interferometer slits. Any information revealing which atom emitted a detected photon reduces the interference visibility~\cite{Wineland93, Ferdinand96, Ketterle2025}. In atomic interferometry, Ramsey fringes provide a closely related manifestation of WP information in the time domain and can be understood within a framework analogous to a Mach–Zehnder interferometer (MZI)~\cite{Bertet2001}. Here, the Ramsey pulses act as the beam splitters of the MZI, giving rise to two interfering paths labeled by the atomic energy state. Any information on the atom's energy reduces the coherence of the atomic dipole and the subsequent contrast of the Ramsey fringes.

\begin{figure*}[t]
\includegraphics[width=\linewidth]{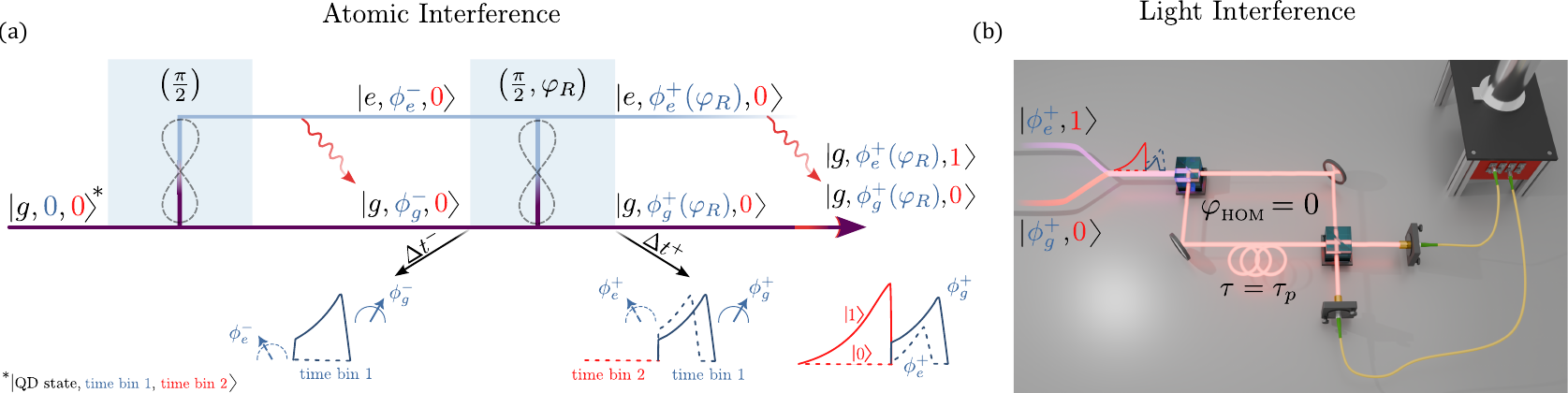}
\caption{Which-Path information in a Ramsey-like sequence. (a) A quantum dot is driven by two classical $\pi/2$-pulses with relative phase $\varphi_R$, separated by a delay $\Delta t$. The photonic field $\phi^{-}_{g (e)}$ emitted in the first time bin $\tau_1$ carries which-path information about the quantum dot state, and alters the coherent emission $\phi^{+}_{g (e)}$ in the second time bin $\tau_2$. In our experiments, Which-path information is accessed immediately before ($\Delta t^-$) and after ($\Delta t^+$) the second Ramsey pulse. 
(b) Which-path information carried by the first time bin at $\Delta t^+$ is probed through the interference visibility in a path-unbalanced Mach-Zehnder interferometer. A delay line matched to the laser repetition period ($\tau_p$) overlaps successively emitted photonic states at the second beam splitter.}\label{fig:1}
\end{figure*}

Here, we exploit both atomic and optical interferometry to access WP information in a Ramsey-like sequence performed on a self-assembled semiconductor quantum dot (QD), see Fig.~\ref{fig:1}. The QD, initially prepared in its ground state, is successively driven by two classical $\pi/2$-pulses (further referred to as Ramsey pulses) with tunable relative phase $\varphi_R$ and delay $\Delta t$. Unlike standard sequences, the delay between pulses can be large enough to allow for spontaneous emission events, such that the QD releases coherent superpositions of zero- and one-photon Fock states in two successive time bins. The spontaneously emitted photonic field emitted in the first time bin correlates with the QD energy state, such that this time bin acts as a WP detector for the Ramsey interferometer, see Fig.~\ref{fig:1}. The extracted WP information increases over time, before being partially erased by the second pulse. We experimentally measure the WP information contained in the first time bin at two different times along the Ramsey sequence: immediately before ($\Delta t^-$) and after ($\Delta t^+$) the second Ramsey pulse. The WP information recorded before the second pulse is accessed through the reduction in the Ramsey fringe contrast, which captures the decreased ability of the QD to coherently absorb light. After the second pulse, the information still present in the first time bin impacts the QD's ability to coherently emit light into the second time bin. We access it through the reduction of the fringe contrast in a self-homodyne interference using a path-unbalanced Mach-Zehnder interferometer (MZI). A simple expression captures the coherent relation between the WP information before and after the second pulse. We experimentally verify this relation, that is allowed by a remarkable agreement between the measurements and the theoretical predictions.\\

\noindent \textit{Which-Path information in the Ramsey interferometer --} In our experiments we use a charged QD acting as an effective two-level atom with ground and excited states $\ket{g}$ and $\ket{e}$, respectively, lowering operator $\hat{\sigma}_- = \ket{g}\bra{e}$, and transition frequency $\omega_0$. The QD is driven through a Ramsey-like sequence, illustrated in Fig.~\ref{fig:1}(a), using excitation pulses with a FWHM much shorter than the radiative lifetime of the QD ($\gamma^{-1}=202.4\pm5.1$~ps). The Ramsey sequence is repeated every $12.3$~ns, and the Ramsey pulse separation is varied over the range $0.08\leq\gamma\Delta t\leq1.68$ (see Supplemental Material for more details on the experiment~\cite{suppl}). The first $\pi/2$-pulse prepares the QD state $\ket{+}=(\ket{e}+\ket{g})/\sqrt{2}$ after which it spontaneously decays, emitting a coherent superposition of vacuum and one-photon Fock state in a first time bin, which plays the role of a quantum meter. In the absence of any other source of decoherence than the spontaneous decay, the joint QD-light field state immediately before the second pulse reads:
\begin{equation}
\ket{\psi(\Delta t^{-})}\equiv\ket{\psi^{-}} = \sqrt{p_e^{-}}\ket{e,\phi_e^-} + \sqrt{1-p_e^{-}} \ket{g,\phi_g^-},
\label{eq:psi_minus}
\end{equation}
where $p_e(\Delta t^-) \equiv p_e^{-}=\exp(-\gamma\Delta t)/2$ is the excited state probability of the QD at time $\Delta t^-$. The photonic states correlated with the QD energy states (the meter's states) read $\ket{\phi_e^-} = \ket{0}$ and $\ket{\phi_g^-} = \frac{\ket{0}+\sqrt{1-2p_e^-}\ket{1}}{\sqrt{2(1-p_e^-)}}$, with $\ket{n}$ ($n=0,1$) denoting the $n$-photon Fock state in the first time bin. During the second pulse, the QD coherently absorbs energy, changing its excited population by:
\begin{equation}\label{eq:abs}
\Delta p \equiv p_e^+ - p_e^- =\frac{1}{2} - p_e^{-} + \sigma_-^- \cos(\varphi_R).
\end{equation}
$p_e^+$ is the QD population immediately after the second Ramsey pulse and $\langle\hat{\sigma}_-(\Delta t^-)\rangle \equiv \sigma_-^-$ is the average QD dipole just before the pulse,
\begin{equation}
 \sigma_-^- = w^{-} \sqrt{p_e^{-}(1-p_e^{-})}. \label{eq:dip-} 
\end{equation}
We have introduced the overlap between the meter's states $w^-$:
\begin{equation}
w^{-} \equiv \braket{\phi_e^{-} | \phi_g^{-}}= \frac{1}{\sqrt{2(1-p_e^{-})}}.
 \label{eq:wp_ideal}
\end{equation}
There are subtle connections between path predictability and fringe visibility \cite{EnglertPRL,SiddiquiPTEP,BeraPRA,DanbiSciRep}, however here we shall loosely define the WP information encoded in the first time bin as $1-w^-$. Consistent with the WP interpretation of Ramsey fringes~\cite{Bertet2001}, the cosine term in Eq.~\eqref{eq:abs} captures a quantum interference between the two evolution paths of the QD, whose contrast $\sigma^-_-$ is altered by the distinguishability $w^-$ of the meter's states. In our experiment, $w^{-}$ depends on the delay $\Delta t$: it decreases from $w^- = 1$ at $\Delta t=0$, corresponding to the absence of WP information, to $w^- = 1/\sqrt{2}$ for $\gamma \Delta t \gg 1$, indicating partial information. Notably, the first time bin cannot carry complete information ($w^{-}=0$). Indeed, the presence of vacuum can arise either from the QD being in $\ket{g}$, or in $\ket{e}$ without yet having decayed. Consequently, $w^-$ never vanishes, even in the limit $\gamma \Delta t \gg 1$.

We experimentally access $\Delta p$ by measuring the light intensity at the end of the Ramsey sequence while varying $\varphi_R$, repeating the measurement for different pulse separations $\Delta t$. The intensity is normalized to the QD emission upon full population inversion -- realized by driving the QD with a single $\pi$-pulse -- and the energy provided by the first $\pi/2$-pulse, $\frac{1}{2}\hbar\omega_0$, is subtracted. 

\begin{figure}[t]
\includegraphics[width=\linewidth]{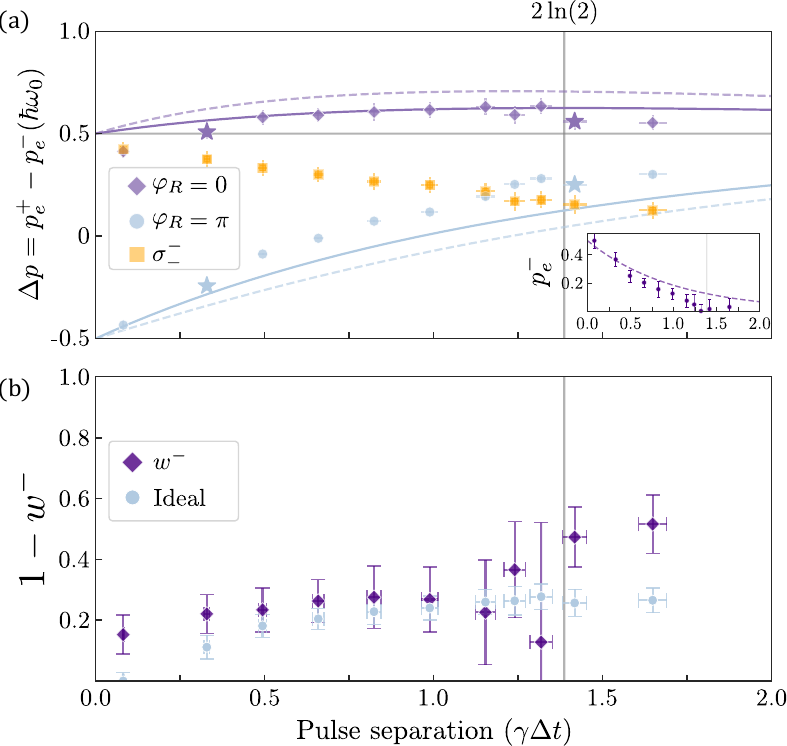}
\caption{(a) The absorption during the second excitation pulse at time $\Delta t$, in units of $\gamma^{-1}$, for a constructive (destructive) Ramsey phase $\varphi=0$ $(\pi)$ in purple diamonds (blue circles), and the atomic dipole $\sigma^-_-$ (fringe visibility) in orange squares. Inset: The population of the qubit right before the second excitation pulse, $\Delta t^-$, as a function of delay.  (b) WP information immediately before the second excitation pulse ($t=\Delta t^{-}$) as a function of the delay. Blue circles show the ideal WP information given by Eq.~\ref{eq:wp_ideal} in the main text, while purple diamonds correspond to the WP information including dipole-induced damping.}\label{fig:2}
\end{figure}

The measured absorption is plotted in Fig.~\ref{fig:2}(a) for Ramsey phases $\varphi_R=0$ and $\pi$ (purple and blue symbols, respectively) together with the theoretical absorption Eq.~\ref{eq:abs} in the presence (solid lines) and the absence ($w^{-}=1$, dashed lines) of WP information. We also display the fringe contrast $\sigma^-_-$ (yellow markers). We observe a good agreement between experimental data and theory in the presence of WP information. For $\gamma\Delta t = 0$, the two $\pi/2$-pulses can act as an effective $\pi$-pulse: depending on $\varphi_R$, the QD is driven to $\ket{e}$ or $\ket{g}$, yielding absorption or emission of $\Delta p = \pm \frac{1}{2}\hbar\omega_0$, respectively. As $\Delta t$ increases, the QD has time to decay between pulses -- see $p_e^-$ as a function of delay in the inset Fig.~\ref{fig:2}(a). This reduces the average QD dipole $\sigma_-^-$ and thus the fringe contrast. Remarkably, for $p_e^{-}<1/2$ (i.e. $\gamma \Delta t>0$) and $\varphi_R=0$, the second pulse yields a larger absorption of energy by the atom than that of a single $\pi/2$-pulse, and is maximal for $\gamma\Delta t=2\ln(2)$. The boost is a purely quantum effect that stems from the increased absorption power of the QD when it is in a coherent superposition of energy states~\cite{Elouard17, Cottet17}. 

We observe a delay-dependent offset with respect to theory which we attribute to spectral diffusion of the QD transition frequency, see Supplemental Material~\cite{suppl}. This spectral wandering exponentially suppresses the Ramsey fringe contrast with increasing delay and shifts the absorption maximum for $\varphi_R=0$ to shorter delays. Nevertheless, at long delays, $\Delta t\gg\gamma^{-1}$, the absorption $\Delta p$ for both phases approaches $\frac{1}{2}\hbar\omega_0$ (not shown here), corresponding to complete decay of the QD prior to the arrival of the second pulse.

Finally, we plot in Fig.~\ref{fig:2}(b) the WP information in the first time bin, $1-w^-$, extracted from the measured values of $\sigma^-_-$ (Eq.~\ref{eq:dip-}, diamond), and compare it to the ideal case where no other decoherence than the spontaneous decay is considered, captured by Eq.~\ref{eq:wp_ideal}. As expected, $1-w^-$ increases over time, signaling the progressive extraction of WP information. The close agreement between the two plots demonstrates that the QD behaves as an ideal two-level system on the timescale of $\gamma^{-1}$, i.e. a few hundred picoseconds. \\

\begin{figure*}[t]
\includegraphics[width=0.8\linewidth]{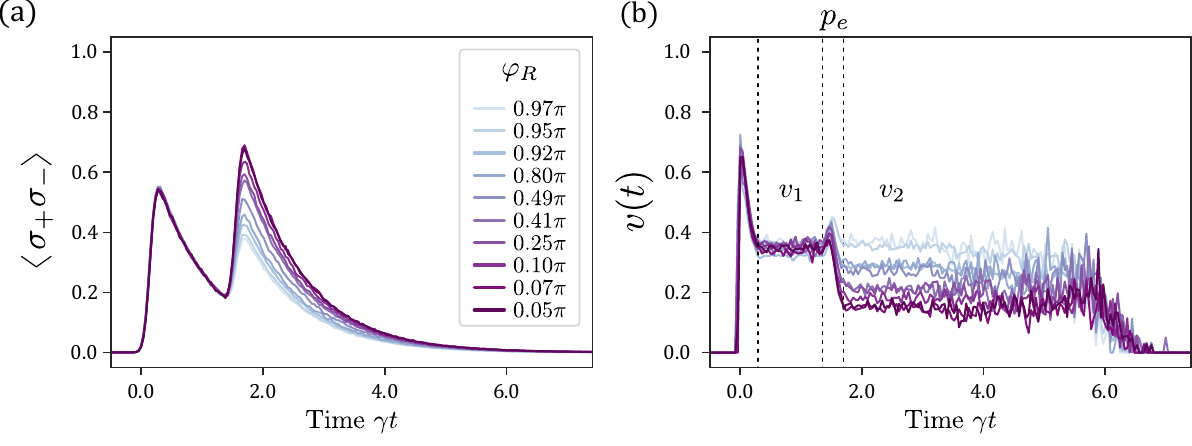}
\caption{\label{fig:3}(a) Time-resolved total energy emission of the QD along the decay profile for various Ramsey phases, at a delay $\gamma\Delta t = 1.42 \pm 0.04$. (b) Time-resolved visibility of the MZI fringes with plateaus $v_{1}$ and $v_{2}$ for the same Ramsey phases and delay as in panel (a). The visibility $v_2$ is reduced due to the which-path information $w^{-}$ encoded in the first time bin (see text).}
\end{figure*}


\noindent \textit{Which-Path Information in Self-Homodyne Interferometry --} 
We now explore how the WP information encoded in the first time bin is modified upon application of the second Ramsey pulse at time $\Delta t$ for two different delays: $\gamma\Delta t \approx 1.42 \pm0.04$ and $\gamma\Delta t\approx 0.33\pm 0.01$ (marked by stars in Fig.~\ref{fig:1}(a)). Although the Ramsey sequence is completed and the QD energy eigenstates no longer explicitly label distinct paths, the photonic field emitted in the first time bin remains correlated with the QD energy states, and thus continues to encode information, which we refer to as WP information such throughout the following. The joint QD-field state immediately after the second pulse, $\Delta t^+$, reads
\begin{equation}
\ket{\psi(\Delta t^{+})}\equiv \ket{\psi^{+}} = \sqrt{p_e^{+}}\ket{e,\phi_e^{+}} + \sqrt{1-p_e^{+}} \ket{g,\phi_g^{+}},
\label{eq:psi_plus}
\end{equation}
where $\ket{\phi_g^{+}}$ and $\ket{\phi_e^{+}}$ denote the new photonic states correlated with the ground and excited states of the QD, respectively, and are given explicitly in the Supplemental Material. The new WP information, $1-w^+\equiv 1- \langle\phi_e^+|\phi_g^+\rangle$, impacts the QD dipole at time $\Delta t^+$ according to
\begin{equation}\label{eq:dip+}
\sigma_-^+ = w^+ \sqrt{p_e^+(1-p_e^+)}.
\end{equation}
$w^-$ and $w^+$ are related through the  expression
\begin{equation}\label{eq:w+w-}
\begin{aligned}
   p_e^+(1-p_e^+) |w^+|^2 = &\left( \frac{1}{2} - p_e^-\right)^2 \\
   &+ p_e^-(1-p_e^-){w^-}^2 \sin^2{(\varphi_R)}.
\end{aligned}
\end{equation}
In our experiments, $w^{\pm}$ depends on $p_e^{\pm}$. To gain further insight, we consider the two quantities as effectively decoupled in two limiting cases: (i) the textbook case of a balanced atomic interferometer, where spontaneous decay is neglected and $p_e^- = 1/2$; and (ii) the limit of two infinitely separated $\pi/2$-pulses, for which $p_e^- = 0$. For (i) we find (see \cite{suppl} for further details):
\begin{equation}\label{eq:wp-balanced}
  |w^{+}| = \frac{w^- \sin{(\varphi_R)}}{\sqrt{1-{w^{-}}^2 \cos^2{(\varphi_R)}}}.
\end{equation}
This expression implies $|w^{+}| \leq w^{-}$: the second pulse can only enhance the distinguishability of the meter's states, with maximal information extraction for $\varphi_R=0$ or $\pi$. Conversely the case (ii) trivially yields $p_e^+=1/2$ and $w^+=1$, i.e. information in the first time bin is completely erased by the second pulse.\\

To measure $w^+$, we exploit the fact that the reduction of the QD dipole captured by Eq.~\eqref{eq:dip+} directly corresponds to a reduction in the spontaneously emitted coherent power $|\braket{\hat{\sigma}}_t|^2$ at time $t$. This coherence is accessed experimentally using self-homodyne measurements performed with an unbalanced MZI, see Fig.~\ref{fig:1}(b)~\cite{Loredo2019, Maillette2023} where two detectors at the output ports (3 and 4) of the final beam splitter record the time-resolved photon numbers $\mu_3$ and $\mu_4$. The time-resolved visibility reads~\cite{suppl}:
\begin{equation}\label{eq:vis}
    v(t) = \frac{|\mu_3(t)-\mu_4(t)|}{\mu_3(t)+\mu_4(t)} = \frac{|\langle\hat{\sigma}_-\rangle_t|^2}{\braket{\hat{\sigma}_+\hat{\sigma}_-}_t}.
\end{equation}
In the ideal case where decoherence stems solely from spontaneous emission, Eq.~\eqref{eq:vis} simply reduces to $v_i = |\langle\hat{\sigma}_-\rangle_{t_i}|^2 / \braket{\hat{\sigma}_+\hat{\sigma}_-}_{t_i}$, where $t_1=0^+$ ($t_2=\Delta t^+$) marks the beginning of the emission in the first (second) time bin. For the Ramsey sequence considered here, the expected visibilities in the first and second time bin are constant and equal to $v_1 = 1/2$ and $v_2 = |w^{+}|^2 (1-p_e^+)$, respectively. Thus the visibility associated to the second time bin provides direct access to the WP information encoded in the first time bin.

\begin{figure*}[t]
\includegraphics[width=0.8\linewidth]{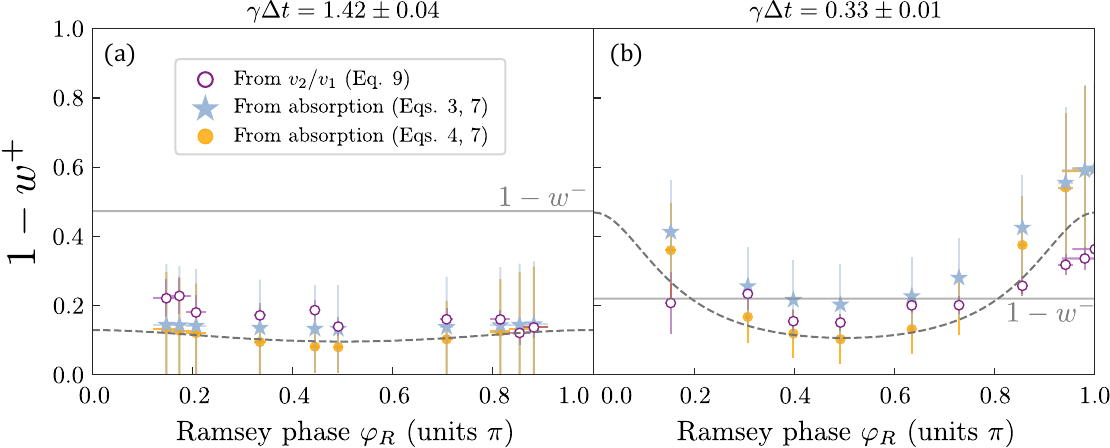}
\caption{(a,b) Modification of the which-path information encoded in the first time bin, as revealed through the coherent emission in the second time bin. Shown is the which-path information $1-w^+$, present in the second time bin $\Delta t^+$ as a function of the Ramsey phase for two different delays $\gamma\Delta t$. The values are obtained either directly from the temporally resolved plateaus (open circles), or inferred from the absorption data using Eq.~\ref{eq:w+w-} assuming ideal $w^-$ (circles) or $w^-$ damped by the presence of a QD dipole (stars). The dashed curves correspond to the theoretical prediction of Eq.~\ref{eq:w+w-}. The solid horizontal line indicates the measured WP $1-w^-$ in the first time bin.\label{fig:4}}
\end{figure*}

Fig.~\ref{fig:3}(a) displays the integrated QD emission intensity resolved along the QD decay profile (i.e. $\braket{\hat{\sigma}_+\hat{\sigma}_-}_t$) for $\gamma\Delta t= 1.42\pm 0.04$ and ten Ramsey phases. Each trace shows two emission peaks, corresponding to the arrival times of the two Ramsey pulses. As the Ramsey phase is tuned from constructive ($0.05\pi$) to destructive ($0.97\pi$) interference, the total emission in the second time bin decreases. Per Ramsey phase, we perform self-homodyne measurements with the emitted field using the unbalanced MZI in Fig.~\ref{fig:1}(b) where one arm contains a delay line matched to the pulse-sequence repetition rate ($81$~MHz). This configuration overlaps successively emitted fields, with same $\varphi_R$, at the final beam splitter. We measure the interference visibility $v(t)$ at the two output ports for a constructive interferometer phase $\varphi_\mathrm{HOM}=0$. 

The time-resolved visibilities for $\gamma\Delta t \approx 1.42\pm 0.04$ ($\gamma\Delta t \approx 0.33\pm 0.01$) are plotted in Fig.~\ref{fig:3}(b). Two sharp peaks in the visibility appear at the arrival times of the excitation pulses. These peaks stem from coherent laser light scattering and are direct evidence of highly ideal emission taking place during excitation. Following each excitation pulse the visibility reaches a constant value: as explained above, this shows that the spontaneous process is ideal -- in particular, it is not affected by pure dephasing. The plateau in the first time bin $v_1$ reaches a value of approximately $0.4$. This is less than expected from a single $\pi/2$-pulse~\cite{Maillette2023}: we attribute this to spin-induced decoherence which is non-negligible on the $12.3$~ns time-scale and uniformly reduces the contrast of the registered visibility over the full photonic wave-packet~\cite{suppl}. This allows us to access $w^+$ through the ratio $v_2/v_1$:
\begin{equation}
    \frac{v_2}{v_1} = 2 |w^+|^2 (1-p_e^+).
\end{equation}
The measured values of $w^+$ as a function of $\varphi_R$ are plotted together with the expected theoretical values in Fig~\ref{fig:4} for two delays. For $\gamma \Delta t\approx1.42$ with $p_e^-=0.03^{+0.06}_{-0.03}$ (see inset Fig.~\ref{fig:2}(a)), $1-w^+$ approaches $0$ signaling the erasure of WP information by the second pulse. Conversely, for $\gamma \Delta t\approx0.33$ and $p_e^- \sim 1/2$, the behavior of $1-w^+$ progressively approaches that of a balanced beam splitter. In this regime, the second pulse can enhance the indistinguishability for extreme values of $\varphi_R$, resulting in a gain in WP information. These behaviors are in qualitative agreement with the analysis above, and in remarkable agreement with the theoretical predictions.\\

\textit{Conclusion and outlook --} We have exploited atomic and optical interferometry to access which-path information and study how it impacts the coherent absorption and emission of light by a QD in a Ramsey sequence. Our measurements are consistent and show remarkable agreement with the theory, proving that QDs can support the exploration of fundamental mechanisms of quantum optics. Our study is a textbook example to the emerging field of quantum energetics \cite{AuffevesElouard2026QuantumEnergetics}, which focuses on how quantum features impact the flows of energy between quantum systems. Our findings point toward the definition of an energy-based entanglement witness, that will be the subject of future work.\\
\nocite{Ollivier2021, Overhauser_Imamoglu}

\textit{Acknowledgments --} This work was done within the Center for Nanosciences and Nanotechnologies micro-nanotechnologies platforms and partly supported by the RENATECH network. A.A. acknowledges the National Research Foundation, Singapore through the National Quantum Office, hosted in A*STAR, under its Centre for Quantum Technologies Funding Initiative (S24Q2d0009), the Plan France 2030 through the projects NISQ2LSQ (Grant ANR-22-PETQ-0006), OQuLus (Grant ANR-22-PETQ-0013), and OECQ, and the ANR Research Collaborative Project “Qu-DICE” (Grant ANR-PRC-CES47). C.A-S. acknowledges the support from the projects from the Ministerio de Ciencia e Innovaci\'on PID2023-148061NB-I00 and PCI2024-153425, the Fundaci\'on Ram\'on Areces, the Grant “Leonardo for researchers in Physics 2023” from Fundaci\'on BBVA, the French Embassy in Spain in the project call ``Appel à projets scientifiques 2022 chercheurs confirmés" and the support from the Comunidad de Madrid fund “Atracci\'on de Talento, Mod. 1”, Ref. 2020-T1/IND-19785. 

\bibliography{references_manual.bib}
\clearpage
\newpage

\onecolumngrid

	\appendix
	\setcounter{figure}{0} \renewcommand{\thefigure}{S.\arabic{figure}}
	\setcounter{equation}{0} 
	\renewcommand{\theequation}{S.\arabic{equation}}
	\setcounter{table}{0} 
	\renewcommand{\thetable}{S.\arabic{table}}

{\begin{center}
    \large{Supplemental Material}
\end{center}}

\subsection{Accessing the which-path information in the first time bin}
For our experiments, we use a charged semiconductor quantum dot, placed inside a cryostation kept at 5~K. We benchmark the performance of our quantum dot by resonantly exciting it with a laser operating at 925~nm. At full population inversion, we measure a single-photon purity of $1-g^{2}(0) = (97.24\pm0.05)\%$, and indistinguishability of $M_s = (93.43\pm 0.39)\%$~\cite{Ollivier2021}. In addition, time-tagged measurements yield a quantum dot lifetime of $\gamma^{-1}=202.4\pm 5.1$~ps.

A Ramsey sequence is implemented using a Michelson interferometer. The excitation laser, operating at a repetition rate of 81~MHz, is sent onto a 50:50 beam splitter, where it is divided into two pulses. One pulse propagates through an arm containing a mirror mounted on a piezoelectric actuator, which allows controlled scanning of the Ramsey phase $\varphi_R$ as a function of the applied voltage. The second arm contains a mirror mounted on a coarse delay stage, enabling precise tuning of the pulse separation $\Delta t$. The two pulses are recombined at the beam splitter and directed onto the quantum dot, where they implement the Ramsey excitation sequence illustrated in Fig.~1(a) in the main text. In the first set of measurements we measure the absorption $\Delta p$ for varying pulse delay $\Delta t$. We continuously scan the Ramsey phase and measure the emission intensity as a function of time. The intensity oscillates as a function of the Ramsey phase, and we extract the fringe maxima ($\varphi_R = 0$) and minima ($\varphi_R = \pi$) and plot them in Fig.~2(a) as a function of pulse separation $\Delta t$. The fringe contrast directly gives us $\sigma^-_-$, and the offset gives us $1/2-p_e^-$. These values are then used to compute the which-path information in the first time bin according to Eqs.~3, 4 in the main text.

\subsection{Spectral diffusion}
The absorption during the second $\pi/2$-pulse is given by 
\begin{equation}\label{eqs:idealp}
\Delta p =\frac{1}{2} - p_e^{-} + \sigma_-^- \cos(\varphi_R),
\end{equation}
where the population right before the pulse is given by $p_e^-=\frac{1}{2}e^{-\gamma\Delta t}$. The observed delay-dependent offset between the experimental data and this ideal expression for $\phi_R =0$ and $\pi$ in Fig.~2(a) arises from spectral diffusion of the quantum dot transition frequency. In between Ramsey pulses, i.e. during $\Delta t$, slow fluctuations in the electrostatic environment cause the quantum dot resonance frequency to wander relative to the fixed laser frequency. This introduces an additional time-dependent phase accumulation $\Delta\omega\cdot \Delta t$ during the Ramsey sequence, where $\Delta\omega$ represents the instantaneous detuning. The modified expression for $\Delta p$ becomes
\begin{equation}
\Delta p =\frac{1}{2} - p_e^{-} + \sigma_-^- \cos(\varphi_R+ \Delta \omega \cdot \Delta t). 
\end{equation}

\begin{figure*}[t]
\includegraphics[width=0.6\linewidth]{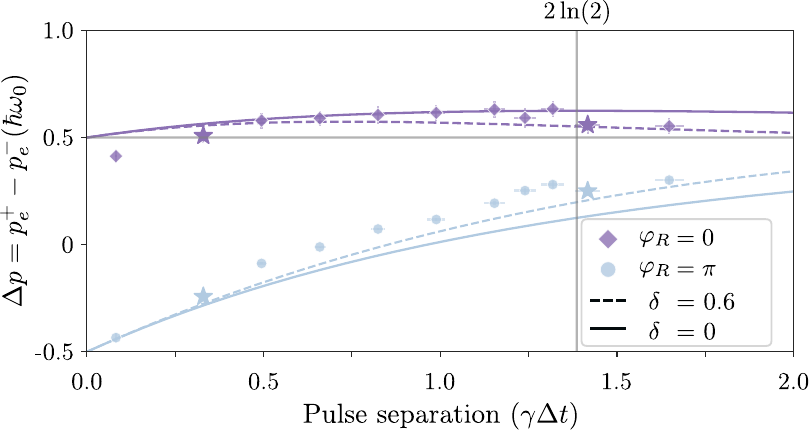}
\caption{The absorption during the second Ramsey pulse as a function of pulse delay with (solid line) and without (dashed line) spectral diffusion $\delta$ of the quantum dot transition frequency. The data shown corresponds to the data in the main text for two different Ramsey phases $\varphi_R$.\label{fig:spectral_diff}}
\end{figure*}

Averaging over the Gaussian distribution of detunings (with effective spectral wandering width $\delta$) over the experimental integration time yields
\begin{equation}\label{eqs:spectral}
    \langle \Delta p \rangle = \frac{1}{2} - p_e^{-} +e^{-\frac{\delta^2 \Delta t^2}{2}} \sigma_-^- \cos(\varphi_R).
\end{equation}
This exponential decay of the Ramsey contrast with $\Delta t$ captures the spectral dephasing, and shifts the expected peak of absorption to delays $\gamma \Delta t < 2\ln(2)$, see Fig.~\ref{fig:spectral_diff} where we plot the ideal theory for $\delta = 0$ (Eq.~\ref{eqs:idealp}) and the theory for $\delta = 0.6$  (Eq.~\ref{eqs:spectral}).

\subsection{Explicit expression of the wavefunction after the second Ramsey pulse }
The second Ramsey pulse implements the unitary map:
\begin{eqnarray}
 \ket{e} & \rightarrow (\ket{e}-e^{-i\varphi_{R}} \ket{g})/\sqrt{2} \\ \nonumber
 \ket{g} & \rightarrow (e^{i\varphi_{R}} \ket{e}+ \ket{g})/\sqrt{2}
 \end{eqnarray}
 
Plugging the map into the wavefunction at time $\Delta t^{-}$ (Eq.~\eqref{eq:psi_minus}), we find the state at time $t=\Delta t^{+}$ reported in Eq.~\eqref{eq:psi_plus}. The pointer states appearing in the wavefunction read: 
\begin{align}
\ket{\phi^{+}_{e}}=e^{i\varphi_R} \frac{ \sqrt{1-e^{-\gamma\Delta t}} \ket{1}+\left(e^{-\gamma\Delta t/2-i\varphi_{R}} + 1\right) \ket{0} }{2\sqrt{p_e^{+}}};\\ \nonumber
\ket{\phi^{+}_{g}}=\frac{\sqrt{1-e^{-\gamma\Delta t}} \ket{1}-\left(e^{-\gamma\Delta t/2-i\varphi_{R}} - 1\right)\ket{0}}{2\sqrt{1-p_e^{+}}};\\ \nonumber
\text{with}~p_{e}^{+}=\frac{1+e^{-\gamma\Delta t/2}\cos{(\varphi_{R})}}{2}=\frac{1}{2}+\sigma_{-}^{-}\cos{(\varphi_{R})}.
\end{align}
Hence, the which-path information at time $\Delta t^{+}$ reads:
\begin{align}
w^{+}\equiv \langle \phi_{e}^{+}\vert \phi_{g}^{+}\rangle= e^{-i\varphi_R}\frac{1-e^{-\gamma\Delta t}+i e^{-\gamma\Delta t/2}\sin{(\varphi_{R})}}{2\sqrt{(1-p_{e}^{+})p_{e}^{+}}},
\end{align}
whose square modulus verifies Eq.~\eqref{eq:w+w-}.
When $\varphi_{R}=0$ or $\varphi_{R}=\pi$, the expression of the which path simplifies reading:
\begin{align}
w^{+}=\pm \sqrt{1-e^{-\gamma\Delta t}},
\end{align}
where $\pm$ correspond respectively to $\varphi_{R}=0$ and $\varphi_{R}=\pi$.
In this regime we have perfect which-path distinguishability ($w^{+}=0$) in the limit $e^{-\gamma\Delta t} \rightarrow 1$, and total indistinguishability ($w^{+}=1$) in the opposite limit $e^{-\gamma\Delta t} \rightarrow 0$.  

\subsection{Time-resolved interference visibility}

To access the which-path information in the second time bin, $\Delta t^+$, we phase-stabilize the Michelson interferometer. This configuration implements a Ramsey sequence in which the Ramsey phase between the two pulses is fixed.
The optical fields emitted following each Ramsey sequence are sent to a Mach-Zehnder interferometer (Fig.~\ref{fig:1}(b)). One arm of the interferometer contains a delay line matched to the laser repetition period, such that optical fields emitted in two consecutive Ramsey sequences (with same Ramsey phase) are temporally overlapped at the second beam splitter. The relative phase of the Mach–Zehnder interferometer is denoted $\varphi_\mathrm{HOM}$. Photons exiting the two output ports of the beam splitter are detected using two superconducting nanowire single-photon detectors (SNSPDs, Single Quantum EOS), and arrival times are recorded with respect to a laser clock reference. The resulting lists of time tags are cut into 500~ms time chunks, a time over which we assume the interferometer phase to be stable.\\

For each detector and each time chunk, we first compute the integrated intensity as a function of time (resolution 500~ms). These intensities exhibit anticorrelated oscillations as a function of $\varphi_\mathrm{HOM}$ between the two output ports due to the presence of photon-number coherence~\cite{Loredo2019, Maillette2023}. We then post-select the data on the interferometer phase $\varphi_\mathrm{HOM}=0$ (equivalent $\varphi_\mathrm{HOM}=\pi$), and reconstruct the temporal profiles at each detector by integrating over the 500~ms windows. This procedure is repeated for several Ramsey phases. 

For each Ramsey phase, we use the temporal profiles to compute the time-resolved visibility, which provides direct access to the which-path information. The visibility is defined as 
\begin{equation}
    v(t)\equiv\frac{|\mu_3(t)-\mu_{4}(t)|}{(\mu_{3}(t)+\mu_{4}(t))},
\end{equation}
where $\mu_{i}$ ($i=1, 2, 3, 4$) denotes the time-resolved number of photons detected at the two input (1,2) and output (3,4) ports of the beam splitter.

Using the fact that two copies of the field sent in the two input ports of the balanced beam splitter are identical, we can rewrite the visibility as $v(t)=|\langle b(t)\rangle|^2/\langle b^{\dagger}(t)b(t)\rangle$ where $b(t)$ is an operator destroying a field's excitation at time t, and verifying the commutation relation $[b(t),b^{\dagger}(t')]=\delta(t-t')$. Using the input-output relations $\langle b(t)\rangle=\sqrt{\gamma}\langle \sigma_{-}\rangle_{t}$ and  $\langle b^{\dagger}(t) b(t)\rangle=\gamma\langle \sigma_{+}\sigma_{-}\rangle_{t}$, we can express the visibility in terms of atomic operators $v(t)=|\langle \sigma_{-}\rangle_{t}|^2/\langle \sigma_{+}\sigma_{-}\rangle_{t}$. \\

This visibility of interference at $\varphi_\mathrm{HOM} = 0 (\pi)$ can be directly related to the which-path information as we show in the following. We will derive an explicit expression for $v(t)$ directly from the average values of field's amplitude and intensity computed using its wavefunction in the long-time limit. Starting from Eq.~\eqref{eq:psi_plus} at time $\Delta t^+$, the QD emits a single photon and decays to the ground state. The light field is then disentangled from the QD and finally its state is given by $\left|\psi\right\rangle = \sqrt{1-p_{e}^{+}}\left|\phi_{g}^{+}\right\rangle \left|0\right\rangle +\sqrt{p_{e}^{+}}\left|\phi_{e}^{+}\right\rangle \left|1\right\rangle$, where $\left|1\right\rangle = \left(\int_{\Delta t}^{\infty} dt \sqrt{\gamma}e^{-\gamma (t-\Delta t)/2} b^{\dagger}(t)\right)\ket{0}$. Writing the state of the field explicitly:
\begin{align}
&\ket{\psi}=\frac{e^{i\varphi_{R}}}{2} \left(\int_{0}^{\Delta t} dt \sqrt{\gamma}e^{-\gamma t/2} b^{\dagger}(t)\right) \left(\int_{\Delta t}^{\infty} dt \sqrt{\gamma}e^{-\gamma (t-\Delta t)/2} b^{\dagger}(t)\right)\ket{0} +\frac{1}{2}\left(\int_{0}^{\Delta t} dt \sqrt{\gamma}e^{-\gamma t/2} b^{\dagger}(t)\right)  \ket{0}\\ \nonumber&+\frac{(e^{i\varphi_{R}}+e^{-\gamma\Delta t/2})}{2} \left(\int_{\Delta t}^{\infty} dt \sqrt{\gamma}e^{-\gamma (t-\Delta t)/2} b^{\dagger}(t)\right)\ket{0}+\frac{(1-e^{-\gamma\Delta t/2-i\varphi_{R}})}{2}\ket{0} ,
\end{align}
where $\ket{0}$ is the vacuum of the propagating field (all possible empty modes).
Using the state above we can compute the amplitude and the intensity profiles of the field, respectively:
\begin{align}
\langle \psi \vert b(t) \vert \psi\rangle= \Theta (\Delta t- t) \frac{\sqrt{\gamma} e^{-\gamma t/2}}{2}
+\Theta (t-\Delta t) \frac{\sqrt{\gamma} e^{-\gamma (t-\Delta t)/2 +i\varphi_{R}}}{2} \left[\frac{1-e^{-\gamma\Delta t}  +(1-e^{-\gamma\Delta t/2+i\varphi_{R}})(1+e^{-\gamma\Delta t/2-i\varphi_{R}})}{2}\right],
\end{align}
and
\begin{align}
\langle\psi \vert b^{\dagger}(t)b(t)\vert \psi\rangle= \Theta (\Delta t- t) \frac{\gamma e^{-\gamma t}}{2} +\Theta (t-\Delta t) \frac{\gamma e^{-\gamma (t-\Delta t)}}{2} \left[\frac{1-e^{-\gamma\Delta t} +|1+e^{-\gamma\Delta t/2-i\varphi_{R}}|^2}{2}\right],
\end{align}
putting together the above expressions we get:
\begin{align}
v(t)=\Theta (\Delta t- t) \frac{1}{2}+\Theta (t-\Delta t)|w^{+}|^2(1-p_{e}^{+}).
\end{align}

\subsection{Spin-decoherence model for the loss of coherence }
As mentioned in the main text, the visibility of the self-homodyne interference is lowered by the decoherence of the electron spin of the quantum dot (QD). The minimal model capable to capture this effect entails a more accurate description of the QD. In this description the QD is modeled as a degenerate four-level system where the two ground states, $\left\{\ket{\uparrow},\ket{\downarrow}\right\}$, belong to an electron and have spin projections $\pm 1/2$ along the quantization axis $z$ (defined by the growth direction of the QD); and the two excited states, $\left\{\ket{ \uparrow \downarrow \Uparrow},\ket{\uparrow \downarrow \Downarrow}\right\}$, belong to an optically excited spin-hole pair, or trion, and have spin projections along $z$ equal to $\pm 3/2$.\\

Conservation of the angular momentum gives the selection rules: the transition $\ket{\downarrow}\rightarrow\ket{\uparrow\downarrow\Downarrow}\left(\text{resp.} \ket{\uparrow}\rightarrow\ket{\uparrow\downarrow\Uparrow}\right)$ is coupled to left- (resp. right-) circularly polarized photons. It is convenient to decouple energy and spin degrees of freedom, adopting the notation: $\left\{\ket{ \uparrow},\ket{\downarrow}\right\}\rightarrow \left\{\ket{\uparrow}\ket{g},\ket{\downarrow}\ket{g}\right\}$ and $\left\{\ket{ \uparrow \downarrow \Uparrow},\ket{\uparrow \downarrow \Downarrow}\right\}\rightarrow \left\{\ket{\uparrow}\ket{e},\ket{\downarrow}\ket{e}\right\}$, hence defining ladder operators acting on spin and energy: $s_{-} \equiv \vert\downarrow\rangle\langle\uparrow\vert,~\sigma_{-}\equiv\vert g\rangle\langle e\vert$. Within this notation, the interaction Hamiltonian (interaction picture) capturing the light-matter coupling between the QD and the electromagnetic field is $V_{I}(t)=i \hbar  \sqrt{\gamma}~\sigma_{-}\left(\vert\downarrow\rangle \langle\downarrow\vert b_{L}^{\dagger}(t) +\vert \uparrow\rangle \langle\uparrow\vert b_{R}^{\dagger}(t)\right)-\mathrm{h.c.}$, where the subscripts $L$ and $R$ of the field's operators stand for left and right circularly polarized photons respectively. The Ramsey pulses have horizontal linear polarization and the QD emission is collected in the vertical linear polarization. Therefore, in the absence of a magnetic field turning on the spin dynamics, the QD can be treated as an effective two-level system as we do in the main text.\\ 

However, even in the absence of an external magnetic field, some spin dynamics can arise due to the electron's hyperfine interaction with the nuclear spins of the $\sim100$ atoms that compose the QD nanostructure, this interaction can be modeled as an effective magnetic field, called Overhauser field, $\textbf{B}^{OH}$, around which the electron's spin effectively precesses~\cite{Overhauser_Imamoglu}. The Hamiltonian coupling between the spin and the Overhauser field can be written as $
H_{\text{s}}=\frac{\Omega}{2} \vert g\rangle\langle g\vert \textbf{n}\cdot \textbf{s}$, with $\Omega/2=g_{\text{el}}\mu_B |\textbf{B}^{OH}|$, where $g_{\text{el}}$ is the Land\'e factor of the electron spin and $\mu_{B}$ the Bohr magneton, $\textbf{n}=\left(\sin{(\theta)}\cos{(\phi)}; \sin{(\theta)}\sin{(\phi)};\cos{(\theta)}\right)$, and $\textbf{s}=(s_x; s_y; s_z)$ being the vector of spin Pauli matrices.

Here, we suppose that the spontaneous emission rate $\gamma$ and the typical Rabi frequency induced by the Overhauser field $\Omega$ verify $\gamma \gg \Omega$, such that the spin dynamics can be completely neglected during the spontaneous emission process. On the other hand, the time $\tau_{p}=12.5$ ns between two successive repetitions of the experiment is comparable with that of the spin dynamics and hence the latter must be included in the description of the self-homodyne interference of two successive photonic wavepackets generated by the double excitation of the trion at times $t=0$ and $t+\tau_p$.

The Overhauser field's magnitude, $|\textbf{B}^{OH}|$, is of the order of tens of mT and its direction, $\textbf{n}$, fluctuates as result of the changing nuclear environment. Over time scales of the duration of a few hundred nanoseconds, the Overhauser field can be considered static such that the spin evolution, for a given configuration, is dictated by the unitary $U(t)=\text{exp}\lbrace -i H_{s} t/\hbar\rbrace$. Then, if the spin is prepared in $\ket{\uparrow}(\ket{\downarrow})$, it evolves in the superposition $C_t \ket{\uparrow} + S_t \ket{\downarrow}(C_{t}^* \ket{\downarrow} +  S_{t}^* \ket{\uparrow})$, where we have introduced the coefficients $C_{t} = \cos\left(\frac{\Omega t}{2}\right) - i \cos{(\theta)} \sin\left(\frac{\Omega t}{2}\right)$ and $S_{t} = \sin{(\theta)}e^{-i\phi} \sin\left(\frac{\Omega t}{2}\right)$,  with $\theta$ and $\phi$ the azimuthal and polar angles of the Overhauser field. Then the quantum mechanical average of the spin polarization $\langle s_{z}\rangle_{t}$ is $(|C_{t}|^2-|S_{t}|^2)$ if the spin's initial state was $\ket{\uparrow}$, and its negative if the spin's initial state was $\ket{\downarrow}$. Due to the ignorance on the Overhauser field configuration, the experimentally observed average of the spin polarization, $\overline{\langle s_z\rangle_t}$, is given by statistical average of $\langle s_{z}\rangle_{t}$ over all possible configurations, i.e. magnitudes and orientations of the field. The average is performed by using an isotropic Gaussian distribution centered around zero and having standard deviation $w$ equal to the decoherence rate of the spin, yielding (for spin initially in $\ket{\uparrow}$):
\begin{align}\label{eq_Merkulov}
\overline{\langle s_z\rangle_t}=\frac{\int_{0}^{\infty} \Omega^2~d\Omega\int_{0}^{\pi} \sin(\theta)~d\theta  \int_{0}^{2\pi} d\phi~e^{-\Omega^2/2w^2}\langle s_z\rangle_{t}}{(2\pi w^2)^{3/2}}=\frac{1}{3}\left( 1 + 2 e^{-w^2 t^2/2}(1 - t^2 w^2)\right).
\end{align}

At the beginning of the sequence the spin is not polarized, its state reads $\rho_s =( \ket{\uparrow}\bra{\uparrow} + \ket{\downarrow}\bra{\downarrow})$. We analyze the counts $\mu_{i} $ registered in the output port, $i\in\lbrace 3,4\rbrace$, of the second beam splitter of the MZI as $\mu_{i} = (\mu_{i,\uparrow} + \mu_{i,\downarrow})/2$, where $\mu_{i,\uparrow} $ ($\mu_{i,\downarrow} $) are the counts obtained if the spin is initially in the state $\ket{\uparrow}$ ($\ket{\downarrow}$). As it will be clear by the end of this section $\mu_{i,\uparrow} = \mu_{i,\downarrow} $, thus, the following calculations are made for the case where the spin is initially in the state $\ket{\uparrow}$ and we will denote $\mu_{i,\uparrow}$ as $\mu_{i}$ to lighten the notation. In this case the experimentally observed spin polarization is given in Eq.~\eqref{eq_Merkulov}, and from now on, we shall denote $\overline{\langle s_z \rangle_{\tau_p}}={\cal C}_{\tau_p}$ to emphasize that this quantity is the correlation between the spin states of two consecutive repetitions of the sequence.\\

The Ramsey pulses are horizontally (H-)polarized and, within our assumption $\gamma\gg\Omega$, the H-polarized laser induces the following map on the QD with obvious notations
\begin{eqnarray}\label{eq_map_separate}
\ket{\uparrow,0}_0 \rightarrow  \ket{\uparrow}   (\sqrt{p} \ket{1_R} + \sqrt{1-p} \ket{0})_0,
\end{eqnarray}
where we have introduced the time tag (subscript) $_0$ that labels the first photonic wavepacket. At time $\tau_p^{-}$ (just before the second excitation sequence), the joint state of the spin and the first wavepacket reads 
$\ket{\Psi(\tau_p^{-})} = (C_{\tau_p} \ket{\uparrow} + S_{\tau_p} \ket{\downarrow} ) \otimes   (\sqrt{p} \ket{1_R} + \sqrt{1-p} \ket{0})_0,$
which becomes at the beginning of the second sequence: 
\begin{eqnarray}
& \ket{\Psi(\tau_p^+)}  = C_{\tau_p} \ket{\uparrow} (\sqrt{p} \ket{1_R} + \sqrt{1-p} \ket{0})_{\tau_p} \otimes (\sqrt{p} \ket{1_R} + \sqrt{1-p} \ket{0})_0 \\ \nonumber
& + S_{\tau_p} \ket{\downarrow} (\sqrt{p} \ket{1_L} + \sqrt{1-p} \ket{0})_{\tau_p} \otimes (\sqrt{p} \ket{1_R} + \sqrt{1-p} \ket{0})_0.
\end{eqnarray}
At this point it is convenient to rewrite $\ket{\Psi(\tau_p^+)}$ in the linear polarization basis using the transformation
\begin{eqnarray}
\ket{1_R} = \frac{\ket{1_H}-i\ket{1_V}}{\sqrt{2}} \\ \nonumber
\ket{1_L} = \frac{\ket{1_H}+i\ket{1_V}}{\sqrt{2}}
\end{eqnarray}
Moreover, sending the state $\ket{\Psi(\tau_p^+)}$ into the MZI maps the wavepacket labeled $0(\tau_p)$ on the input port $1$ ($2$) of the second beam splitter. We finally rewrite this input state in the following way
\begin{align}\label{eq_inputBS}
 \ket{\Psi_{in}}  & = C_{\tau_p} \ket{\uparrow}  \ket{\psi_{+,1}} \otimes  \ket{\psi_{+,2}} + S_{\tau_p} \ket{\downarrow}  \ket{\psi_{+,1}} \otimes  \ket{\psi_{-,2}}\\  \nonumber
 \text{with}~\ket{\psi_{+,{j}}}  &= (-i\sqrt{p/2} \ket{1_{V}} + \sqrt{1-p} \ket{0} +\sqrt{p/2} \ket{1_{H}})_{j}\\  \nonumber
 \ket{\psi_{-,{j}}} &= (i\sqrt{p/2} \ket{1_{V}} + \sqrt{1-p} \ket{0} +\sqrt{p/2} \ket{1_{H}})_{j},
\end{align}
where $j \in \{ 1, 2 \}$ denotes the BS input ports. Now acknowledging that the detection is along the vertical polarization (V) (crossed-polarized detection scheme), we introduce the operators $b_{V,1}(t)$ and $b_{V,2}(t)$ corresponding respectively to the first and second input ports of the BS.  Considering that the BS is balanced and introduces a phase $\varphi$ we can write the expressions of the counts $\mu_{i}$ registered at the output ports $i\in\{3,4\}$:
\begin{align}
\mu_3 &= \langle b_{V,1}^{\dagger} b_{V,1} \rangle +\langle b_{V,2}^{\dagger} b_{V,2} \rangle + 2\text{Re}\langle e^{-i\varphi} b_{V,1} ^{\dagger} b_{V,2} \rangle \\ \nonumber
\mu_4 &= \langle b_{V,1}^{\dagger} b_{V,1} \rangle +\langle b_{V,2}^{\dagger} b_{V,2} \rangle - 2\text{Re}\langle e^{-i\varphi} b_{V,1} ^{\dagger} b_{V,2} \rangle 
\end{align}
where the average is taken on the state $\ket{\Psi_{in}}$ of Eq.~\eqref{eq_inputBS}. We simply have $\langle b_{V,1}^{\dagger} b_{V,1} \rangle = \langle b_{V,2}^{\dagger} b_{V,2} \rangle = p/2$ while the cross terms read:
\begin{eqnarray}
\langle  b_{V,1} ^\dagger b_{V,2} \rangle= |C_{\tau_p}|^2 \bra{\psi_{+,1},\psi_{+,2}} b_{V,1}^\dagger b_{V,2} \ket{\psi_{+,1},\psi_{+,2}} + |S_{\tau_p}|^2 \bra{\psi_{+,1},\psi_{-,2}} b_{V,1}^\dagger b_{V,2} \ket{\psi_{+,1},\psi_{-,2}}
\end{eqnarray}
We can then estimate
\begin{eqnarray}
&\bra{\psi_{+,1}} b_{V,1} \ket{\psi_{+,1}} = \bra{\psi_{+,2}} b_{V,2} \ket{\psi_{+,2}} = i \sqrt{\frac{p}{2}}\sqrt{1-p} \\ \nonumber
&\bra{\psi_{-,2}} b_{V,2} \ket{\psi_{-,2}} = - i \sqrt{\frac{p}{2}}\sqrt{1-p},
\end{eqnarray}
and finally get
\begin{eqnarray}
\langle  b_{V,1} ^\dagger b_{V,2} \rangle = (|C_{\tau_p}|^2 - |S_{\tau_p}|^2) \frac{p}{2}(1-p).
 \end{eqnarray}
The visibility of the fringes reads $v= \text{Max}_{\varphi}|\mu_{3}-\mu_{4}|/(\mu_{3}+\mu_{4})= 2|\langle  b_{V,1} ^\dagger b_{V,2} \rangle|/p$, yielding after statistical averaging over the Overhauser field's configurations:
\begin{eqnarray}\label{eq_bare_visibility}
\overline{v}& = \vert(\overline{|C_{\tau_p}|^2} - \overline{|S_{\tau_p}|^2})\vert (1-p) = |{\cal C}_{\tau_p}| (1-p).
 \end{eqnarray}
Since, as mentioned before, taking $\ket{\downarrow}$ as spin's initial state would only flip the sign of ${\cal C}_{\tau_p}$, the visibility would be the same.

We conclude that the loss of visibility is due to the factor $|{\cal C}_{\tau_p}|$ whose expression is given by Eq.~\eqref{eq_Merkulov} with $t=\tau_p$. Considering that when $w \tau_p\approx 0.5$, $|{\cal C}_{\tau_p}|\approx 0.7$ that is approximately the ratio between the measured plateaus and their corresponding expected values, we can infer that the spin coherence lifetime during the experiment was $w^{-1}\approx25$ ns. 

\subsection{Relation between WP information before and after the second Ramsey pulse}

The QD dipole before and after the second Ramsey pulse are related by the simple equation
\begin{equation}
    |\sigma^+_-|^2 = \left( \frac{1}{2} - p_e^-\right)^2 + |\sigma^-_-|^2 \sin^2{(\varphi_R)},
\end{equation}
which is equivalent to Eq.~\eqref{eq:w+w-} in the main text:
\begin{equation}
   p_e^+(1-p_e^+) |w^+|^2 = \left( \frac{1}{2} - p_e^-\right)^2 + p_e^-(1-p_e^-){w^-}^2 \sin^2{(\varphi_R)}.
\end{equation}
Conversely, we recall that $p_e^+$ and  $p_e^-$ are related through Eq.~\eqref{eq:abs} in the main text, yielding
\begin{equation}
   p_e^+ =  \frac{1}{2} + \sqrt{p_e^-(1-p_e^-)} w^- \cos{(\varphi_R)}.
\end{equation}
If $p_e^-=1/2$ (balanced interferometer), we find $p_e^-(1-p_e^-)=1/4$ and $p_e^+=(1+w^-\cos(\phi_R))/2$, hence $p_e^+(1-p_e^+)=(1-{w^-}^2\cos^2(\phi_R))/4$, and finally
\begin{equation}
  |w^{+}| = \frac{w^- \sin{(\varphi_R)}}{\sqrt{1-{w^{-}}^2 \cos^2{(\varphi_R)}}}.
\end{equation}
Conversely, if $p_e^-=0$ then $p_e^+=1/2$ and $|w^+|=1$.

\end{document}